\documentclass[twoside]{dis07}
\usepackage[latin1]{inputenc}
\usepackage[dvips]{graphicx,epsfig,color}
\usepackage{wrapfig,rotating}
\usepackage{amssymb,amsmath,array}

\begin{document}
\title{Generalized Parton Distributions from Hadronic Observables}

%***********************************************************************
% AUTHORS INFORMATION AREA
%***********************************************************************
\author{S. Ahmad$^1$, H. Honkanen$^1$, S. Liuti$^1$ and S.K. Taneja$^2$
%
% Optional short acknowledgment: remove next line if non-needed
%\thanks{This is an optional funding source acknowledgment.}
%
% DO NOT MODIFY THE FOLLOWING '\vspace' ARGUMENT
\vspace{.3cm}\\
%
% Addresses and institutions (remove "1- " in case of a single institution)
1- University of Virginia - Physics Department \\
382, McCormick Rd., Charlottesville, Virginia 22904 - USA
%
% Remove the next three lines in case of a single institution
\vspace{.1cm}\\
2- Ecole Polytechnique, \\ CPHT, F91128 Palaiseau Cedex, France
\\
\\
}
%***********************************************************************
% END OF AUTHORS INFORMATION AREA
%***********************************************************************

\maketitle

\begin{abstract}
We propose a physically motivated parametrization for the unpolarized 
generalized parton distributions, $H$ and $E$, valid at both zero and non-zero values
of the skewness variable, $\zeta$. 
At $\zeta=0$, $H$ and $E$ are determined using constraints from simultaneous fits 
of experimental data on both  
the nucleon elastic form factors and the deep inelastic structure functions.
Lattice calculations of the higher moments
constrain the parametrization at $\zeta > 0$.
Our method provides a step towards a model independent extraction 
of generalized distributions from the data that is
alternative to the mathematical ansatz of 
double distributions.
\end{abstract}

\section{Introduction}
Generalized Parton Distributions (GPDs) parametrize the soft  
contributions in a variety of hard exclusive 
processes, from Deeply Virtual Compton Scattering (DVCS)
to hard exclusive meson production (see \cite{Die_rev,BelRad} for reviews).
The conceptual idea behind their definition 
allows one to address a vast, previously inaccessible 
phenomenology, from the simultaneous description of  hadronic structure  
in terms of transverse spatial and
longitudinal momentum degrees of freedom \cite{Bur}, to   
the the access to the description of angular momentum of partons in nucleons and nuclei 
via Ji's sum rule \cite{Ji1}.

At present, a central issue is the definition of a quantitative, reliable 
approach beyond the construction of GPDs from specific models and/or particular 
limiting cases, that can incorporate new incoming experimental data 
in a variety of ranges of the scale $Q^2$, and the four-momentum transfer between the incoming
and outgoing protons, $\Delta \equiv (t,\zeta)$. 
The matching between measured quantities and Perturbative QCD (PQCD) based predictions   
for DVCS should proceed, owing to specific factorization theorems, 
similarly to the extraction of Parton Distribution Functions (PDFs) from deep 
inelastic scattering. 
A few important caveats are however present since
GPDs describe {\em amplitudes} and are therefore more elusive observables
in experimental measurements.
Experiments delivering sufficiently accurate data have, in fact, just begun \cite{halla}. 
The comparison with experiment and the formulation of parametrizations 
necessarily encompasses, therefore, other strategies using additional constraints, 
other than from a direct comparison with the data.     

We propose a strategy using a combination of experimental data on nucleon 
form factors, PDFs, and lattice calculations of Mellin moments with $n \geq 1$. The latter,
parametrized in terms of Generalized Form Factors (GFFs), were  
calculated by both the QCDSF \cite{QCDSF_1} 
and LHPC \cite{LHPC_1}
collaborations for both the unpolarized and polarized 
cases up to $n=3$, therefore allowing to access the skewness dependence of GPDs.

\begin{wrapfigure}{r}{0.5\columnwidth}
\centerline{\includegraphics[width=0.45\columnwidth]%
{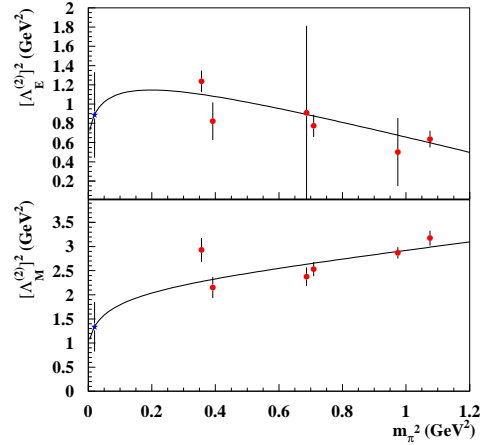}}
\caption{(color online) The dipole masses squared for $n=2$, for the isovector magnetic 
(lower panel) and electric (upper panel) contributions obtained by performing fits to the 
lattice results of \protect\cite{QCDSF_1}. 
The value at the physical
pion mass obtained from our fit is also shown (star).}\label{fig1}
\end{wrapfigure}

\section{Generalized Parton Distributions from Lattice Moments}
GPDs can be extracted most cleanly from DVCS \cite{Ji1}.   
In this contribution we concentrate on the unpolarized scattering
GPDs, $H$, and $E$, from the vector ($\gamma_\mu$)
and tensor ($\sigma_{\mu\nu}$) interactions, respectively. 
We adopt the following set of kinematical variables:
$(\zeta, X, t)$, where $\zeta= Q^2/2(Pq)$
is the longitudinal momentum transfer between the initial 
and final protons ($\zeta \approx x_{Bj}$ in the asymptotic limit,
with Bjorken $x_{Bj} = Q^2/2M\nu$),
$X=(kq)/(Pq)$ is the momentum fraction relative to the initial proton carried by the struck 
parton, $t = -\Delta^2$, is the four-momentum transfer squared. 
%%%
$X$ is not directly observable, it appears in the amplitude as an integration variable
\cite{Die_rev,BelRad}. The need to deal with a more complicated phase space,  
in addition to the fact that DVCS interferes coherently with the Bethe-Heitler (BH) process,  
are in essence the reasons why it is more challenging to extract GPDs from experiment, 
wherefore guidance from phenomenologically motivated parametrizations becomes important. 

\noindent
We first present a parametrization of $H$ and $E$ in the flavor Non Singlet (NS) sector,
valid in the $X>\zeta$ region, obtained by extending our 
previous zero skewness form \cite{AHLT1}, through proper kinematical shifts:
\begin{equation}
H(X,\zeta,t)  =  G_{M_{X}}^{\lambda}(X,\zeta,t) \, R(X,\zeta,t) 
\label{param1_H}
\end{equation}
%%%%
(a similar form was used for $E(X,\zeta,t)$),
where $R(X,\zeta,t)$ is a Regge motivated term  
describing the low $X$ and $t$ behaviors, while
$G_{M_{X}}^{\lambda}(X,\zeta,t)$, was obtained using a spectator model.

In order to model the  $X < \zeta$ region, we observe that 
the higher moments of GPDs give 
$\zeta$-dependent constraints, in addition to the ones from the nucleon form factors. 
The $n=1,2,3$ moments of the NS combinations: $H^{u-d} = H^u-H^d$, and $E^{u-d} = E^u-E^d$ 
are available
from lattice QCD \cite{QCDSF_1,LHPC_1}. 
They can be written in terms of the isovector components as: 
\begin{eqnarray}
\label{HuHdn}
H_n^{u-d} \equiv \int \, dX X^{n-1} (H^u - H^d) & = & \frac{\tau (H_M^V)_n + (H_E^V)_n}{1+\tau} \\
\label{EuEdn}
E_n^{u-d} \equiv \int \, dX X^{n-1} (E^u - E^d) & = &  \frac{(E_M^V)_n - (E_E^V)_n}{1+ \tau},
\end{eqnarray}
where the l.h.s. quantities are obtained from the lattice moments 
calculations, whereas $(H_{M(E)}^V)_n$ and $(E_{M(E)}^V)_n$ are amenable to chiral 
extrapolations.
We used lattice calculations for the unpolarized GFFs
obtained by the QCDSF collaboration using two flavors of ${\mathcal O}(a)$-improved 
dynamical fermions for several values of $t$
in the interval 
$0 \lesssim t \lesssim 5$ GeV$^2$, and covering a range of pion mass values, 
$m_\pi \gtrsim 500 \, {\rm MeV}^2$.
Similarly to previous evaluations \cite{LHPC_1} 
the GFFs for both $H$ and $E$, display a dipole type behavior for all three $n$ values,
the value of the dipole mass increasing with $n$. 
We performed an extrapolation by extending 
to the $n=2,3$ moments a simple ansatz proposed in \cite{Ash} for the nucleon form factors 
that: {\it i)} uses the connection between
the dipole mass and the nucleons radius; {\it ii)} introduces a modification of the non analytic 
terms in
the standard chiral extrapolation that suppresses the contribution of chiral loops at large $m_\pi$.
Despite its simplicity, the ansatz seems to reproduce both the lattice results 
trend at large $m_\pi$ 
while satisfying the main physical criteria {\it i)} and {\it ii)}.
Our results for the dipole mass at $n=2$ are shown in Fig.\ref{fig1}.

%%%%%%%%%% FIGURE 2
\begin{wrapfigure}{r}{0.5\columnwidth}
\centerline{\includegraphics[width=0.45\columnwidth]%
{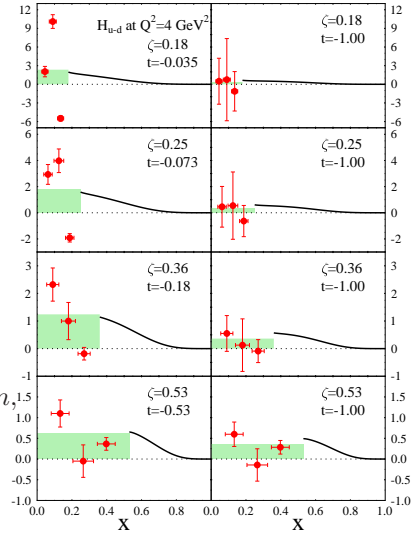}}
\caption{(color online) Comparison of $H^{u-d}$ for 
different values of $\zeta$% = 0.18, 0.25, 0.36, 0.53$, 
and  
$-t \equiv t_{min}= 0.035, 0.073, 0.18, 0.53$ GeV$^2$ (left panel), and $-t = -1$ GeV$^2$ 
(right panel), 
calculated using the procedure described in the text.}
\label{fig2}
\end{wrapfigure}
%%%%%%%%%%%%%%%%%%%
\section{Reconstruction from Bernstein Polynomials}
Similarly to the PDFs case \cite{Yndurain}, 
with a finite number of moments in hand, one can use 
reconstruction methods attaining weighted averages of the 
GPDs, around average ranges of $X$. The weights are provided by 
the complete set of Bernstein polynomials.
%Through method we are able to provide
%an intuitive 
%insight on the behavior of the  function, even if not a point-wise description.  

In Fig.\ref{fig2} we show $H^{u-d}$ reconstructed using the available 
lattice moments. We performed the procedure in the $X<\zeta$ 
region only using:
\begin{equation}
\overline{H}_{k,n}(\zeta,t) = 
\int\limits_0^\zeta H(X,\zeta,t) b_{k,n}(X,\zeta) dX \; \; \; k=0,...n, 
\label{berns1}
\end{equation}
where:
$b_{k,n}(X,\zeta) = X^k \, (\zeta-X)^{n-k}/
\int\limits_0^\zeta X^k \, (\zeta-X)^{n-k} \, dX$, and we used subtracted moments, defined
as: 
\begin{eqnarray}
\label{missing_a}
\left( H_{n} \right)_{X<\zeta}  = H_{n} - \int\limits_\zeta^1 \,  H^I(X,\zeta,t) X^{n} dX  , 
\end{eqnarray}   
\noindent
%\vspace{0.5cm}
where  $H_{n}$ are the Mellin moments, and 
$H^I(X,\zeta,t)$ was obtained from Eq.(\ref{param1_H}). 
For $n=2$, $k=0,1,2$, the reconstruction procedure  
yields \cite{AHLT2}:
%\begin{subequations}
\begin{eqnarray}
\label{berns_zeta}
\overline{H}_{02}( \zeta X_{02})& = & \frac{1}{\zeta^3} \left\{ 3 A_{10}^\zeta \, \zeta ^2 
- 6 A_{20}^\zeta \, \zeta  + 3 
\left[A_{30}^\zeta + \left( -\frac{2\zeta}{2-\zeta} \right)^2 A_{32} \right] \right\},
\nonumber \\
\overline{H}_{12}(\zeta X_{12}) & = & \frac{1}{\zeta^3} \left\{ 6 A_{20}^\zeta \, \zeta - 6 
\left[ A_{30}^\zeta +  \left( -\frac{2\zeta}{2-\zeta} \right)^2 A_{32} \right] \right\},
\nonumber \\   
\overline{H}_{22}(\zeta X_{22}) & = & \frac{1}{\zeta^3} \left\{ 3 A_{30} + 
\left( -\frac{2\zeta}{2-\zeta} \right)^2 A_{32}  \right\} , 
\end{eqnarray}
%\end{subequations}
%%%
where $X_{01}=0.25$, $X_{02}=0.5$, $X_{03}=0.75$, and $A_{10}, A_{20}, A_{30}, A_{32}$ are the
GFFs from Ref.\cite{QCDSF_1}.   

In conclusion, we provided a fully quantitative parametrization
of the NS GPDs, valid in the region of Jefferson Lab experiments \cite{halla} 
that, differently from model calculations, and for the first time to our knowledge, 
makes use of experimental data in combination with lattice results. 
Given the paucity of current direct experimental measurements of GPDs, 
our goal is to provide more stringent, model independent  
predictions that will be useful both for model builders, in order 
to understand the dynamics of GPDs, and for the planning of future 
hard exclusive scattering experiments.

\section{Acknowledgments}

We thank Ph. Haegler, P. Kroll and G. Schierholz for useful comments. We are also grateful to 
J. Zanotti for providing us with the recent lattice 
calculations from the QCDSF collaboration.
This work is supported by the U.S. Department
of Energy grant no. DE-FG02-01ER41200 and NSF grant no.0426971. 

% ****************************************************************************
% BIBLIOGRAPHY AREA
% ****************************************************************************

\begin{footnotesize}
% IF YOU DO NOT USE BIBTEX, USE THE FOLLOWING SAMPLE SCHEME FOR THE REFERENCES
% ----------------------------------------------------------------------------

% ----------------------------------------------------------------------------

\end{footnotesize}

% ****************************************************************************
% END OF BIBLIOGRAPHY AREA
% ****************************************************************************

\end{document}